\documentclass[conference, 10pt]{IEEEtran}
\makeatletter
\def\ps@headings{%
\def\@oddhead{\mbox{}\scriptsize\rightmark \hfil \thepage}%
\def\@evenhead{\scriptsize\thepage \hfil \leftmark\mbox{}}%
\def\@oddfoot{}%
\def\@evenfoot{}}
\makeatother
\usepackage{setspace}
\usepackage{graphicx}
\usepackage{setspace}
\usepackage{epsfig,cite,amsmath,amssymb,latexsym,subfigure}
\usepackage{floatflt}
\usepackage{theorem}
\usepackage{url}
\usepackage{times}
\usepackage[usenames]{color}
\newcommand{\Prob}[1]{\mathcal{P}\left(#1\right)}
\newcommand{\set}[1]{\left[#1\right]}

\newcommand{\eqnref}[1]{(\ref{eqn:#1})}

\newcommand{\eqnlabel}[1]{\label{eqn:#1}}

\newcommand     {\paren}[1]{\left(#1\right)}

\newcommand{\curlb}[1]{\left\{#1\right\}}

\newcommand{\eX}[1]{\e^{#1}}

\newcommand{\e}{e}

\title{ \vspace{-0.3cm}\bf \large The Quality of Source Location Protection in Globally Attacked Sensor Networks}

\author{
\normalsize Silvija Kokalj-Filipovi\'c, Fabrice Le Fessant, and Predrag Spasojevi\'c,\\
\small INRIA Saclay, WINLAB Rutgers University,\\
\small\em \{silvija.kokalj-filipovic,fabrice.le\_fessant\}@inria.fr, spasojev@winlab.rutgers.edu
}
\begin{document}
\date{} \maketitle

\begin{abstract}
We propose an efficient scheme for generating fake network traffic to disguise the real event notification in the presence of a global eavesdropper, which is especially relevant for the quality of service in delay-intolerant applications monitoring rare and spatially sparse events, and deployed as large wireless sensor networks with single data collector. The efficiency of the scheme that provides statistical source anonymity is  achieved by partitioning network nodes randomly into several node groups. Members of the same group collectively emulate both temporal and spatial distribution of the event. Under such dummy-traffic framework of the source anonymity protection, we aim to better model the global eavesdropper, especially her way of using statistical tests to detect the real event, and to present the quality of the location protection  as relative to the adversary's strength. In addition, our approach aims to reduce the per-event work spent to generate the fake traffic while, most importantly,  providing a guaranteed  latency in reporting the event. The latency is controlled by decoupling the routing from the fake-traffic schedule. We believe that the proposed source anonymity protection strategy, and the quality evaluation framework, are well justified by the abundance of the applications that monitor a rare event with known temporal statistics, and uniform spatial distribution.
\end{abstract} 
\vspace{-0.2cm} 
\section{Introduction}\label{sec:intro}
\vspace{-0.2cm} 
Privacy issues are an important aspect of monitoring applications in wireless sensor networks {\em (WSNs)}. A recent survey of state-of-the-art research on privacy protection in WSNs \cite{LiSurvey}, among other problems, reviews strategies to protect the 
object observed by a WSN node, referred to as {\em source}, from the global eavesdropper {\em (Eve)} \cite{Mehta07}, which can infer the location of the object based on the established location of the source. 
\vspace{-0.2cm}
\subsection{Problem Description}\label{subsec:problem}
\vspace{-0.2cm}
The observed object may be a smuggler crossing the border, an important person entering a classified area, or endangered animals monitored in their habitats. Messages from the source are
propagated in a traditional hop-by-hop manner, and directed to a fixed data collector, referred to as a base station, or a {\em sink}. Sink protection is not an issue as the adversary
usually knows the sink, in fact she may even know the whole topology of
the sensor network. In addition, Eve detects the timing and location of all transmissions in the network  (hence {\em global}); she can hear and capture
any packet sent in the network (either with a very powerful antenna or
she has her own sensor network deployed in the area). Eve is powerful: she can employ complex statistical algorithms for detection, and arbitrary localization techniques. However, the message itself is
encrypted and Eve does not know the encryption, hence she cannot capture the
message and infer the object's position from the content. Moreover, Eve, despite being so powerful and omnipresent, needs to stay invisible. We define {\em outage} as the event when, following an eavesdropped transmission, Eve reveals itself by taking actions based on the false suspicion that an event occurred. The actions may involve physical presence of the attacker or her faculties, in order to capture or destroy the object. Hence, a false-alarm presents a risk of personal exposure and liability.

The adversary  gathers the source of the transmission based on the
change in the traffic pattern; a conspicuous case would be when a node starts transmitting after a prolonged period of inactivity in the WSN. For many event-reporting applications, despite the fact that the attacker cannot learn the details from the message content,  inferring the {\em contextual} information, i.e. whether, when and where a concerned event has happened, may be enough to jeopardize monitored resources. 
Intuitively, the persistent {\em dummy (fake)
traffic} is the only way to obfuscate the events, and the formal proof for it is available in  \cite{Mehta07}. Dummy packets follow
a predefined schedule, aligned with the expected timeline of real packets, so
that Eve cannot observe the change. To better explain the intricacies of this approach, especially in light of the existing research, we next introduce two models of monitored phenomena. Let us first define the application delay as the delay in event reporting. In both scenarios, the duration of time is relative to the application delay constraint, which is a known value $\Delta$. We assume that the WSN is divided
into cells, such that each sensor node monitors a unique cell
and that the events are occurring in a uniform manner over time and space. An active event is any event that is not reported to the sink yet.
\subsubsection{Scenario $\mathcal{A}$: Frequent and Dense Events} This model describes monitoring of a physical phenomena that creates on average one event per cell over one {\em cycle}, a relatively short period of time, whose duration is larger but of the same order of magnitude as the application latency constraint.  In other words, in any cycle, there are many active events in the network. 
\subsubsection{Scenario $\mathcal{B}$: Rare and Spatially Sparse Events} 
In this model, the events are rare and isolated.  For example, if the allowed application delay is in minutes, the expected interval between  events is measured in hours or days. They are spatially sparse: we assume that there is at most {\em one active event} at a time. In fact, the examples of monitoring applications at the beginning of \ref{subsec:problem} all represent the scenario $\mathcal{B}.$ No single node can statistically emulate the spatial and temporal characteristics of the events in this model. In addition, by observing a single node for the duration several orders of magnitude larger than $\Delta,$ Eve can not reliably deduce deviation from the expected behavior.
Consequently, Eve attempts to observe abnormalities in the network-wide traffic pattern. The anonymity protection scheme described here implements the traffic pattern in a decentralized  manner, so that the occurrence of real events does not cause observable abnormalities.
\subsection{Solution Outline}
The uniform spatial/time distribution of events guides naturally the {\em baseline scenario} for the dummy traffic:
 all cells in the network  send dummy messages at a constant rate regardless of whether  a real event has occurred or not. That means that an event would have to wait to be reported on average for half of the inter-transmission interval. However, since the traffic in the network always keeps the same pattern, it effectively defeats any traffic analysis techniques. The main problem with dummy traffic is immediately obvious from the basic scenario: limiting the reporting delay calls for a high-rate fake traffic, which is not only expensive but may quickly burn out the network.

Our approach stipulates the importance of knowing the event's temporal dynamics, but only in terms of the two introduced scenarios; it allows us to design energy-efficient protection strategies. It is natural to assume that the expected frequency of events will be known to both the attacker and the network architect, given that we design the monitoring application for a particular physical phenomenon (an important person does not enter a classified area every single minute, or even hour), and that the easiest characterization of a random process is through its first moment, or the moment's estimate. Next, among the all-positive probability distributions with a given expected value, exponential distribution has the highest entropy. Hence, assuming that inter-event times follow an exponential distribution of the estimated mean leads to a good and justifiable model of the event's randomness. In terms of traffic overhead/ energy-consumption and interference, the optimal design would force each node to transmit as rarely as possible, and that would be in exponentially distributed intervals of the expected duration {\em exactly} equal to the expected time between real events; smaller intervals generate more traffic and therefore cost more, while larger ones create too few opportunities for embedding the real-traffic, especially under delay constraints. 

The second major underpinning of our approach is the network-centric view of the problem. The following aspects of the problem are looked at from both the event's and the network's perspective:
\begin{itemize}
\item Event is characterized as a spatio-temporal process over the whole network area,
\item Event-reporting delay includes routing latency,
\item Fake-traffic shaping is a decentralized process, collaboratively maintained by all nodes
\item Energy consumption per-event of the protection strategy is equally split among network nodes, and substantially decreased with respect to the baseline strategy due to nodes' collaboration.
\end{itemize}

Finally, our source-anonymity protection scheme aims to achieve statistical event unobservability. The absolute protection under baseline strategy is not applicable to delay-sensitive applications, such is the majority of event monitoring in WSNs. Secondly, we do not adjust the timing of real events as in \cite{ShaoINFOCOM08}, to make the event pass the statistical test under the test parameters assumed to be used by Eve. Instead, we make the event pass the test with the same probability as the dummy transmissions, making it statistically indistinguishable.
\vspace{-0.2cm}
\subsection{Existing Research}
\vspace{-0.2cm}
There are several papers that study the WSN source anonymity but, in our opinion, propose solutions that are efficient under scenario $\mathcal{A}$ only\cite{Mehta07,YangWiSec08,ShaoINFOCOM08}. 
At the expense of substantial traffic overhead, a practical tradeoff between security and latency is proposed in \cite{ShaoINFOCOM08}.
This approach, as ours, assumes that the attacker knows the defense strategy of the WSN, and that she will use a state-of-the-art statistical test to distinguish the real event from the fake once. The rationale is that fake traffic, even if random, is designed to follow a distribution,  while the real events may not.  According to \cite{ShaoINFOCOM08}, the event sources should run the same test, and adjust the time of the real event to pass the test. The event sources test and correct the intervals between {\em their own} transmissions of fake and real events, striving to maintain their exponentiality. The paper compares two varieties of this approach: one in which the real event embeds itself by waiting until next scheduled transmission (ProbRate), and another one (FitProbRate), when it waits as little as needed to pass the goodness-of-fit test for the exponential distribution, inferred from the previous transmissions. The latter approach results in smaller delay, but requires the correction of the schedule, quite likely for all the post-event transmissions. 

We further observe that the scheme proposed in \cite{ShaoINFOCOM08} defines delay as the time between the event occurrence and the source's transmission, which {\em holds only} for WSN applications in which the sink is one hop away from any source. If the packet is delivered to the sink in a hop-by-hop manner, the latency includes another random part due to summation of the exponentially distributed delays associated with such transmission schedule of each relay. We refer to this additional delay as the {\em publishing route (PR) latency}. When the expected value $\lambda$ of the inter-transmission times is the same at each node, as in \cite{ShaoINFOCOM08}, and designed to imitate a relatively rare event pattern, for the source-sink route of $h$ hops, the PR latency becomes an Erlang-distributed random variable with mean $h\lambda.$ From this point of view too, Scenario $\mathcal{B}$ requires a modified approach to source anonymity. For the cases simulated in \cite{ShaoINFOCOM08} , the mean of dummy message intervals is
$20s,$ and real events arrive according to a Poisson process
with the rate changing from $1/20$ to $1/100.$ Their protection scheme achieves  the average latency of less
than 1s. If we replace seconds with  hours, having in mind events that happen once a day, or once a couple of days, the delay of one hour does not seem to be acceptable. Additionally, the PR latency for rare events is prohibitive, even for applications that are not delay-sensitive. 
To decrease the overhead of the dummy protection scheme in \cite{ShaoINFOCOM08}, in \cite{YangWiSec08} the same authors propose a WSN with several proxy nodes, which pick up
transmissions from surrounding nodes, and filter out the dummy packets. Apart from requiring mitigating solutions, frequent dummy traffic inevitably leads to interference, which additionally increases the PR latency. 

The next section briefly introduces our solution to this problem. Section \ref{sec:Decentral} describes and analyzes two decentralized algorithms that implement the proposed solution: one is suitable for uniform spatio-temporal distribution of events, while another is more resilient to the distribution outliers (spatial clusters, and temporal bursts of events), however, more expensive in terms of overhead. Section \ref{sec:BurstOutage} motivates and presents some of the simulations, while some are left out due to space constraints. Finally, in \ref{sec:Conclude}, we conclude.
\vspace{-0.2cm} 
\section{Our Approach}\label{sec:Appr}
\vspace{-0.2cm}
\subsection{System Model}\label{subsec:Model}
We have a static WSN of $n$ nodes. There is one static sink collecting event notifications from all nodes. The monitoring application is delay sensitive: the time between the event occurrence and the sink's notification must be smaller than $\Delta$. We assume that monitored events have Poisson temporal distribution of a known rate $\lambda=1/\mu$, and uniform spatial distribution over the area of network deployment. Hence, the time between the events is distributed according to a exponential distribution $\zeta_{\mu}$ (of expected value $\mu>>\Delta$).
The source is assumed to transmit a burst of packets, all describing the event. Due to space limitations, we here analyze the burst of unit length (one packet), although our second algorithm is designed for exponential distributions perturbed by both random outliers and event bursts. 
\vspace{-0.1cm}
\subsection{Problem Formulation}\label{subsec:Formul}
\begin{figure}[!t] 
\begin{center}
\includegraphics[width=5.0in]{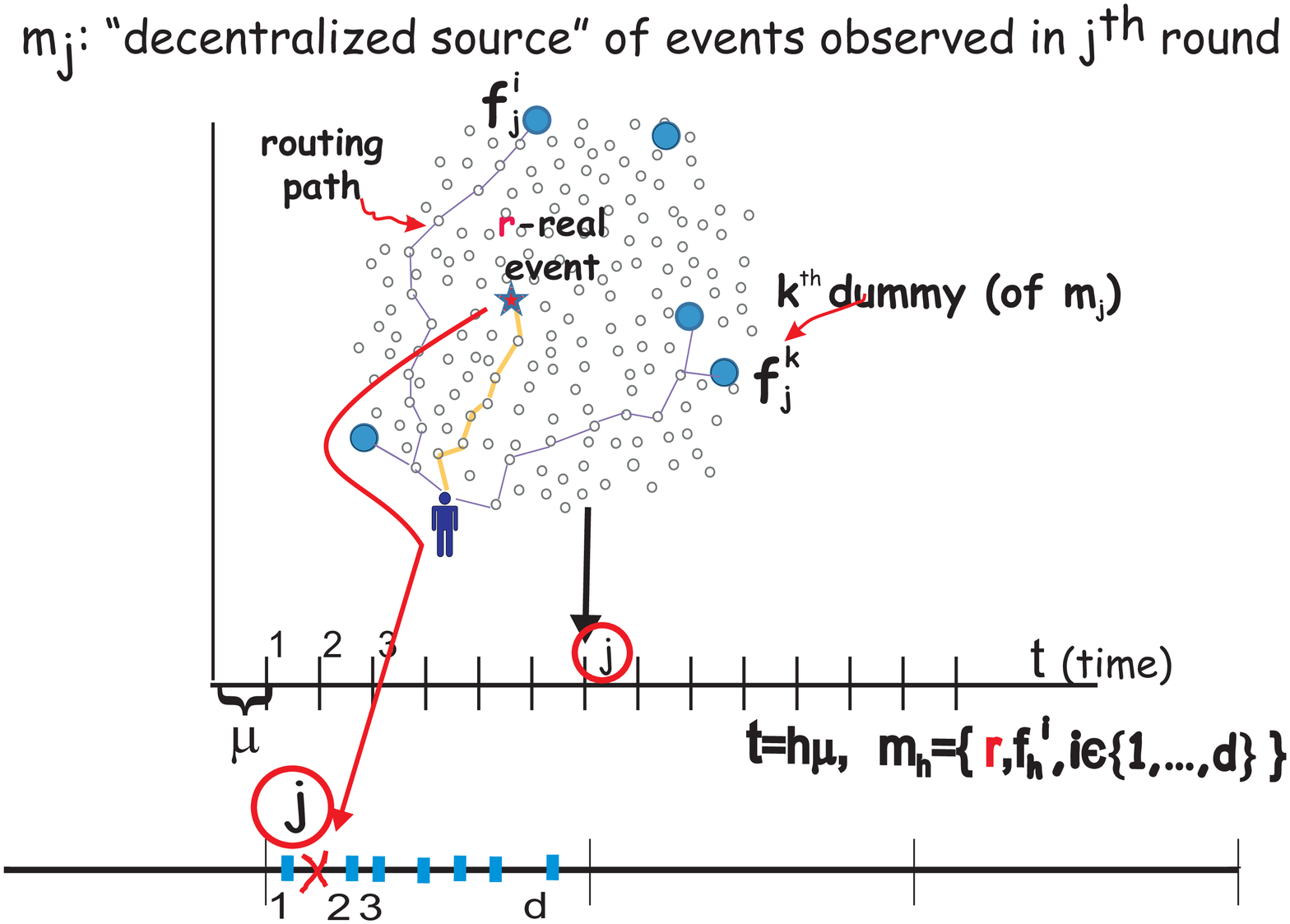}
\vspace{-0.6cm}
\caption{Dummy population of size $d$ accompanies an event; nodes have been assigned the round index $j$ at the initialization.}
\vspace{-0.9cm}
\label{fig:spatial}
\end{center}
\end{figure}

We established that the only way to confuse Eve 
is through persistent network-wide transmissions. Simultaneously, as hop-by-hop is prevalent data transfer model in WSNs, and distance to a sink may be considerable, to satisfy the application latency constraints, we need to decouple routing from the fake traffic schedule by allowing immediate relaying of event notifications as opposed to piggybacking them on the existing fake transmissions. However, as such a route may be backtraced to the source, a similar routing path should be emulated from each fake source (see Figure~\ref{fig:spatial}). 
Given the mentioned constraints, our goal is to achieve statistically strong source anonymity through methods that optimize energy and delay \cite{Mehta07}. We hereby propose a pattern of fake traffic that scales well with network size, and satisfies application-latency constraints, while protecting the source location up to a given  significance level, defined under the strong statistical tests available to Eve \cite{AndersonDarling,WaldSeqAnal}. 
To confuse Eve, we propose to replicate the spatio-temporal process through the following mechanisms:
\vspace{-0.2cm}
\begin{itemize}
\item {\bf Source Emulation:} a subset of $d$ nodes regularly wakes up to act as dummy sources. As a result, any real event is {\em covered} by $d$ dummy sources,  which we refer to as {\em dummy population}. To explain what it means for an event to be covered, we introduce a time interval, dubbed {\em round}, whose duration is equal to the expected inter-event time. Hence, the length of a round is $\mu.$ Covering an event implies the expected existence of fake transmissions in the same round in which the event occurs. To engage all nodes equally, we may divide the network in $d$ groups and assign one representative of the group to a distinct round. Each group will maintain a schedule that emulates the event distribution $\zeta_{\mu}.$ Due to the size of the dummy population, the probability distribution of the intervals between any two consecutive dummy events is $\zeta_{\mu/d}$ (exponential of expected value $\mu/d$). Such cooperative shaping of the fake traffic in order to emulate a sufficiently dense Poisson distribution is amenable to distributed implementation, which is thoroughly explained in Section~\ref{sec:Decentral}. However, as the attacker overhears every transmission that occurs in the network, and may integrate all recorded temporal data into one global network activity timeline, it is judicious to ignore for a moment the decentralized implementation. Instead, we look at the global timeline as if it was produced by a single source sampling the values of event inter-arrivals from the distribution $\zeta_{\mu/d}.$  The joint empirical distribution of inter-transmission times, created by extending the fake schedule with the immediate (undelayed) transmissions of real events, is, based on the transmissions' independence, $\zeta_{\mu/(d+1)},$ i.e. exponential with the expected value $\mu/(d+1).$ For sufficiently large $d,$ $\zeta_{\mu/(d+1)}$ does not diverge perceptibly from the distribution on the global timeline of fake events $\zeta_{\mu/d)}$. 
\item {\bf Route Emulation:} each source (dummy or real) forwards the packet along a predetermined route towards the sink (see Figure~\ref{fig:spatial}). The inter-transmission time between relays is constant and significantly shorter than $\Delta$ (and, consequently, orders of magnitude smaller than $\mu,$ as opposed to \cite{ShaoINFOCOM08} where it is tied to the inter-transmission time of dummy sources). As the real source starts transmitting without delay, the application latency is equal to the routing delay, which is now decoupled from the time dynamics of the fake traffic, and can be further optimized by minimizing the number of hops. 
\item {\bf Knowing the Attacker's Detection Methods:}  Eve is assumed to be able to estimate 
the distribution of recorded transmission times. The estimated distribution is used as a reference point in the real-event detection strategy that involves a statistical test. We base our analysis on the assumption of a single statistical test, namely {\em Anderson-Darling (A-D)} test for exponentiality \cite{AndersonDarling}. Apart from the motives of simplicity, and the existence of a reference that employs the same test \cite{ShaoINFOCOM08}, the additional arguments can be found in \cite{StephensEDF}. The A-D test belongs to the class of {\em goodness-of-fit} tests that evaluate the distance between the distribution of the
sample data and a specified probability distribution. Alternatively, the test evaluates if the current sample comes from the same distribution as the previously evaluated ones. If the
distance is statistically significant, where the significance level is derived from a parameter of the test, which also defines the percentage FA of {\em false alarms}, it is decided that data do not follow this
distribution. 
\end{itemize}
As the latency issue is decoupled from the fake traffic design, we seek to determine the minimal dummy population size needed to secure a given statistical anonymity, hence minimizing the overhead. For a WSN of size $n,$ we define $W_n$  as per-event and per-round energy consumption of the source anonymity mechanism. We express $W_n$ in terms of the number of packet-forwarding hops, where we upper-bound the length of the publishing route by $h=O\paren{\sqrt{\frac{n}{\log{n}}}}$ \cite{GuptaKumarPower}. Hence, the cost of each fake source transmission will be of the same order. With $d$ fake sources covering each real event, 
$W_n=O\paren{(d+1)\sqrt{\frac{n}{\log{n}}}},$ which demonstrates the importance of optimizing $d$, as the source anonymity calls for a sufficiently large $d.$ However, note that $d$ would have to be on the order of $\sqrt{n\log{n}}$ to exert the same overhead as the method in \cite{ShaoINFOCOM08}.
\vspace{-0.2cm}
\section{ Decentralized Generation of Fake Traffic}\label{sec:Decentral}
\vspace{-0.1cm}
\subsection{ Baseline Decentralized Algorithm}
In  Section~\ref{sec:Appr} we highlighted the importance of cooperative and distributed shaping of the fake traffic that should result in sufficiently dense Poisson distribution of dummy transmissions. Let us now propose the realization of such a decentralized system. 
We first establish the baseline for the decentralized implementation. As the dummy population covering each event needs to include on average $d$ fake sources, we define en epoch of duration $T=\mu\frac{n}{d}$ in which each network node will get to be a dummy once. Then, we let each node draw a time instant to transmit in this epoch by sampling uniform distribution $U(0, T).$ The causality of transmissions will arrange all node samples in increasing order, resulting in exponential distribution $\zeta_{\mu/d}$ of inter-transmission times. The procedure can be extended to the consecutive epochs, so that in the $i$th epoch nodes draw their transmission times from $U((i-1)T, iT).$
Note that the particular uniform distribution range does not overlap with the ranges of distributions pertaining to other epochs. The collective empirical distribution of transmission times is the distribution of {\em almost} independent disjoint events, and therefore it approximates the Poissonian distribution. The independence of transmission events is broken only on the boundary of the epochs, as with each new epoch the nodes sample from a uniform distribution of different disjoint range. Hence, the distribution of the interval $Z$ between the first event in the new epoch and the last event in the previous epoch is not exponential.
\begin{align}
\nonumber Z &= T+U-V\\
&= T + \min\curlb{X_1,\cdots,X_n}-\max\curlb{X_1,\cdots,X_n}.\eqnlabel{firstz}
\end{align}
\begin{align}
\nonumber F_U(u) &= \Prob{U\leq u}= 1-\Prob{U>u}\\
\nonumber &= 1-\prod^n_{i=1}{\Prob{X_i>u}}\\
\nonumber &= 1-\prod^n_{i=1}{\paren{1-\Prob{X_i\leq u}}}\\
&= 1-\prod^n_{i=1}{\paren{1-F_X(u)}}=1-\paren{1-F_X(u)}^n.
\end{align}
The probability distribution for $U$ is
\begin{eqnarray}
\nonumber f_U(u) &=& n\paren{1-F_X(u)}^{n-1}f_X(u)\\
&=&\frac{d}{\mu}\paren{1-\frac{u}{T}}^{n-1},\ \ \mbox{for  $T \geq u\geq 0,$}
\end{eqnarray}
and o.w. $0,$ which is for sufficiently large $n$ clearly exponential distribution of expected value $\frac{\mu}{d}$
\begin{eqnarray}
f_U(u) &\approx& \frac{1}{\frac{\mu}{d}}\eX{-\frac{u}{\frac{\mu}{d}}}.
\end{eqnarray}
Further,
\begin{eqnarray}
\nonumber F_V(v) &=& \Prob{V\leq v}= \prod^n_{i=1}{\Prob{X_i\leq v}}\\
&=& \prod^n_{i=1}{F_X(v)}=F^n_X(v).\eqnlabel{cumV}
\end{eqnarray}
\begin{eqnarray}
f_V(v) &=& nF^{n-1}_X(v)f_X(v)=\frac{d}{\mu}\paren{\frac{v}{T}}^{n-1},
\end{eqnarray}
for  $T \geq v\geq 0.$
For sufficiently large $n$
\begin{align}
f_V(v) &=\frac{d}{\mu}\eX{-\frac{T-v}{\frac{\mu}{d}}}.
\end{align}
For $W=T-V,$ $f_W(w) =\frac{d}{\mu}\eX{-\frac{w}{\frac{\mu}{d}}},$ 
where now
\begin{align}
Z&=U+W. \eqnlabel{secondz}
\end{align}
%
The random variable $Z$ has a range $\set{0,2T},$ and, from \eqnref{secondz}, its probability distribution is Erlang, with the shape parameter of $2,$ denoted $\epsilon_{(2,\frac{\mu}{d})}$:
\begin{align}
\nonumber f_Z(z) &= \int^{T}_{0}{f_U(\tau)f_W(z-\tau)d\tau}\\
&= \int^{z}_{0}{\frac{d}{\mu}\eX{-\frac{\tau}{\frac{\mu}{d}}}\frac{d}{\mu}\eX{-\frac{z-\tau}{\frac{\mu}{d}}}d\tau}= z\frac{d^2}{\mu^2}\eX{-\frac{z}{\frac{\mu}{d}}}.
\end{align}
As the probability of an inter-epoch sample in the collection of test samples of size $t,$ where $t<n,$ is $\frac{1}{t}$ or $0,$  the average test-failure probability will be at most $\frac{1}{t}FA_{E}+\frac{t-1}{t}FA,$ where $FA_{E}$ denotes failure probability when Erlang-distributed samples are tested on exponentiality. Even though the distribution of $Z$ differs from $\zeta_{\mu/d},$ 
 we observe that the baseline decentralized implementation will pass the exponentiality test for any reasonable size of $t$ (no estimate is made based on 3 samples). Such a distribution satisfies our needs for real-event obfuscation. However, for other reasons, related to cases when the outliers of the real-event temporal distribution coincide with the spatial correlation of events, we propose the following realization of the dummy traffic, dubbed {\em group implementation}.
\vspace{-0.2cm}
\subsection{ Group Algorithm}
In the initialization, the WSN nodes are divided into $d$ groups of size $\left\lfloor n/d \right\rfloor$, and  every node in the group is assigned an index $i$, denoting the round in which the node will cover the source. Hence, the $k$th round, where $k\equiv i\ \mbox{\cal{mod}}\left\lfloor n/d \right\rfloor,\ i\in\curlb{1,\left\lfloor n/d \right\rfloor},$ will have a dummy population of $d$ nodes belonging to different groups. A group schedule is created by letting the $i$th member select the specific time instant to transmit a fake message, sampled from the uniform distribution $U((i-1)\mu, i\mu)$. Such an algorithm  is amenable to distributed implementation, since each node can independently measure time and keep count of the current round. Once the round index corresponds to node's index modulo group size, the node draws a sample from the pertaining  distribution, and determines its transmission time. As the independence of transmission events is broken only on the boundary of the rounds, the distribution of the interval $Z$ between the first event in the new round and the last event in the previous round is not exponential. Once again, $Z = U+W,$ where $W=\mu -V,$ $V=\max\curlb{X_1,\cdots,X_d},$ and $U=\min\curlb{X_1,\cdots,X_d}.$
\begin{align}
F_U(u) &=1-\paren{1-F_X(u)}^d,\ \ \mbox{and}
\end{align}
\begin{align}
f_U(u) &=\frac{d}{\mu}\paren{1-\frac{u}{\mu}}^{d-1},\ \ \mbox{for  $\mu \geq u\geq 0,$}
\end{align}
and o.w. $0,$ which is, for large enough $d,$ exponential distribution of expected value $\frac{\mu}{d}.$
Now, following a derivation similar to \eqnref{cumV}, we obtain
\begin{align}
f_V(v)&=\frac{d}{\mu}\paren{\frac{v}{\mu}}^{d-1},
\end{align}
for  $\mu \geq v\geq 0.$ For large $d,$ the distribution of $W$ is $\zeta_{\mu/d}.$
The random variable $Z$ has a range $\set{0,2\mu},$ and the probability distribution is
\begin{align}
f_Z(z) &= z\frac{d^2}{\mu^2}\eX{-\frac{z}{\frac{\mu}{d}}}=\epsilon_{(2,\frac{\mu}{d})},
\end{align}
for sufficiently large $d.$ 
%
Hence, with group implementation $Z$ follows the same Erlang distribution $\epsilon_{(2,\frac{\mu}{d})}$ as in the baseline decentralized implementation. However, to be able to state that the test sample of size $t<n,$ which includes inter-round intervals with probability $1/d,$ will be statistically indistinguishable from the sample of exponentially distributed intervals, we need to impose a stricter requirement for the value of $d.$ The average test-failure probability will be $\frac{1}{d}FA_{E}+\frac{d-1}{d}FA.$ Upper-bounding the failure probability $FA_{E}$ (when Erlang samples are tested on exponentiality) with one, we obtain that $d$ should be at least $\frac{1}{FA}$. If the spatially-uniform events do follow the distribution $\zeta_{\mu}$ in time, this may unnecessarily increase the per-event overhead with respect to the baseline.
However, if we have a more complex event distribution, by selecting the group algorithm we are able to not only render the event's temporal characteristics indistinguishable, but also to obfuscate spatial correlation. 
\begin{figure}[!t] 
\begin{center}
\includegraphics[width=3.5in]{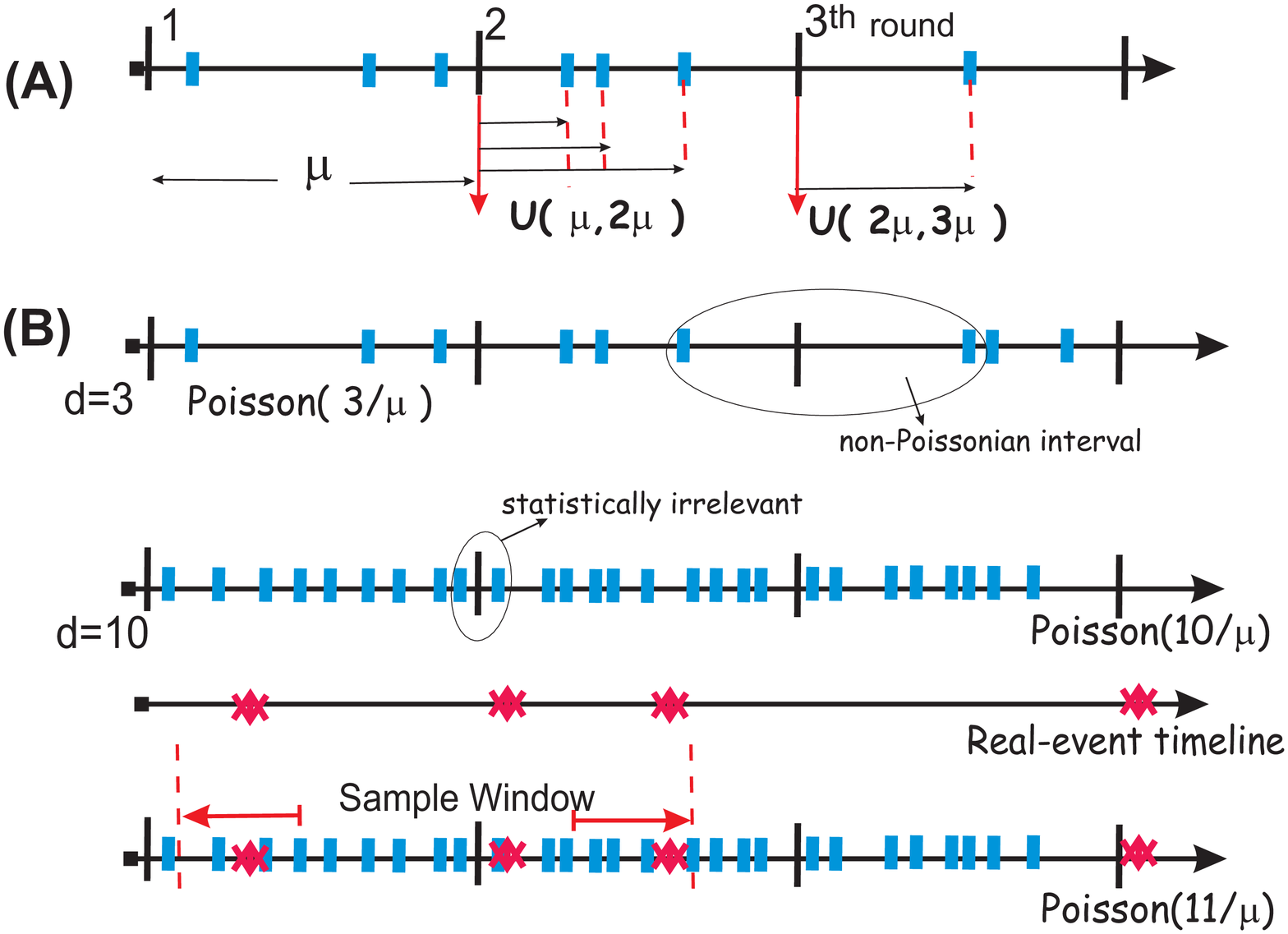}
\vspace{-0.8cm}
\caption{{\bf(A) }Illustration of our group algorithm for fake-traffic generation, where we randomly form $d$ groups of nodes to cooperatively create the schedule by sampling a series of uniform distributions. {\bf(B) }Large enough $d$ renders small divergence from the Poisson distribution statistically irrelevant. Bottom axis: for sufficiently large $d,$ mixing several "group" schedules and the timeline of the real events (red stars) into a global transmission schedule observed by Eve, statistically indistinguishable from Poisson schedule.}
\vspace{-0.8cm}
\label{fig:fitting}
\end{center}
\end{figure}
\vspace{-0.1cm}
\section{Simulations}\label{sec:BurstOutage}
\vspace{-0.2cm}
When designing the simulations, we dismiss the possibility that Eve would test the schedule of any single node, since, in our scenario, inter-transmission times per-node are large with respect to $\mu$ (inter-event times), and, hence, it takes a lot of time to record a reasonable test sample. Our primary goal was to demonstrate the influence of the dummy population size $d$ to the statistical properties of the network-wide transmission schedule, both in the absence of real events, and under different  stochastic models for real events.
For sufficiently large $d,$ which is still much smaller than $n,$ our simulations show that the insertion of events does not statistically change the time axis. Therefore, by running the statistical tests, Eve does not obtain any additional information that would help her capture the monitored object, even if the time of transmission of a real source is not delayed. With the existing work \cite{ShaoINFOCOM08}, an adjustment delay is added, and another mechanism may be needed to fix the sample mean affected by the adjustments, to delude Eve's sequential analysis tests, such as SPRT \cite{WaldSeqAnal}.

Performing the A-D test, which is a powerful test for exponentiality \cite{ADPoisson,GFitExpStephens}, on the samples drawn from an {\em exponential distribution} results in a percentage of failures, which represent false alarms. The percentage of false alarms is a random variable whose mean corresponds to the false-alarm  parameter of the test (also referred to as the {\em significance level}), denoted by $\alpha$. Due to randomness, over certain sets of test samples this percentage will fluctuate around the value of the parameter provided by the test. Our testing strategy monitors the rate of test failures to evaluate if it behaves as expected for the exponential distribution. 
\begin{figure}[t] 
\begin{center}
\begin{tabular}{c}
\epsfig{figure=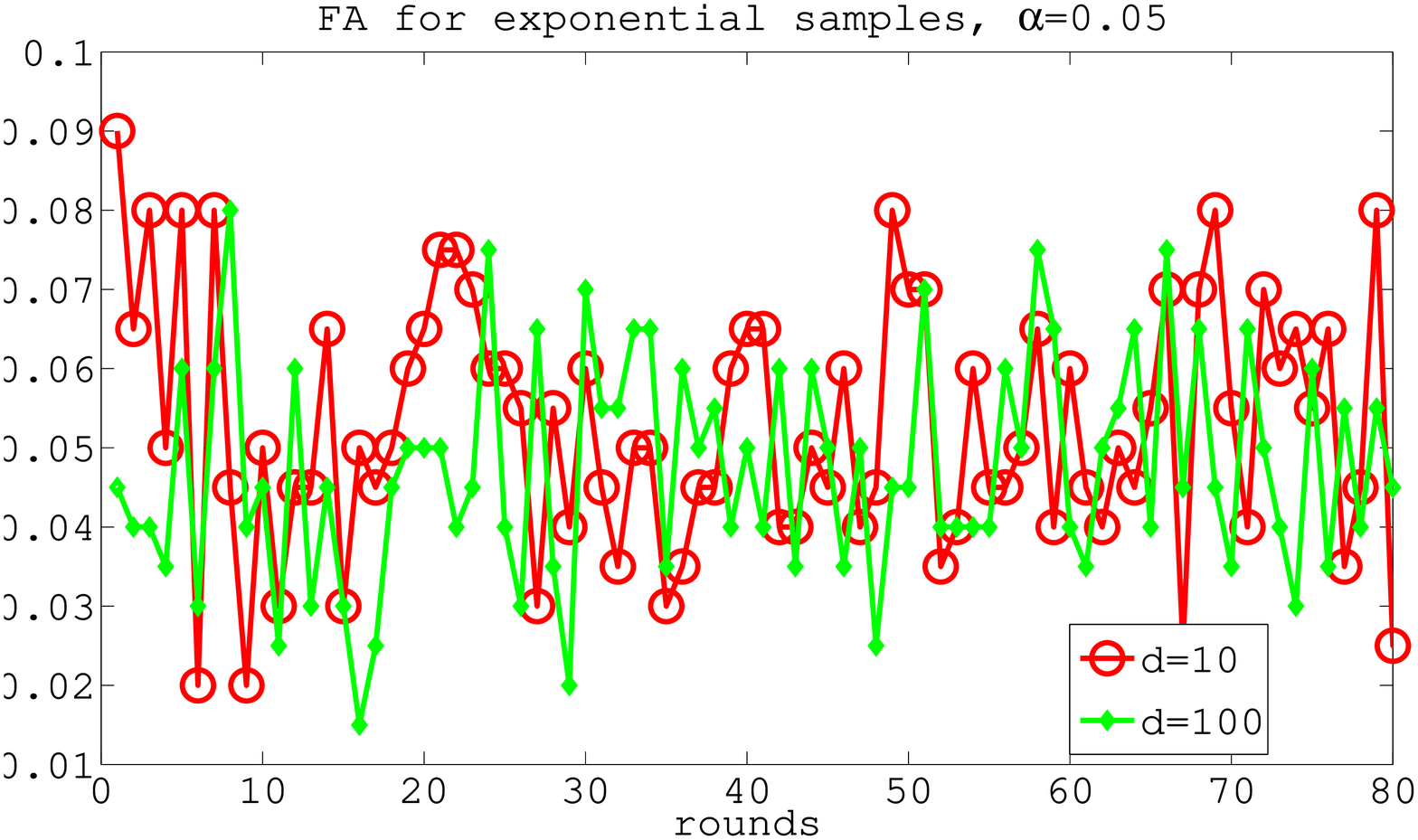, width=2.5in}\\
{\bf (A) - FA has constant mean}\\
\epsfig{figure=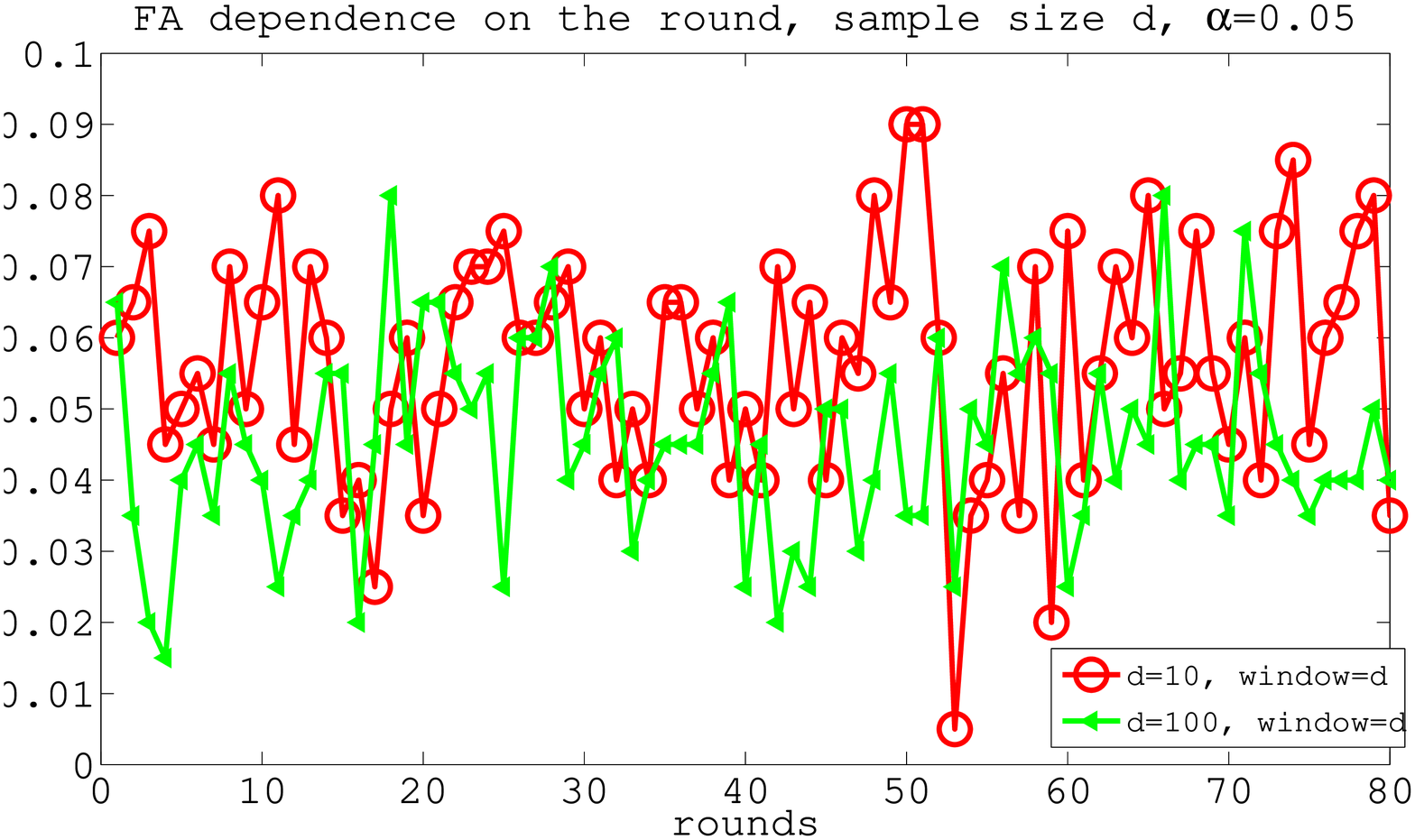, width=2.5in}\\
{\bf (B) - FA has constant mean}\\
\epsfig{figure=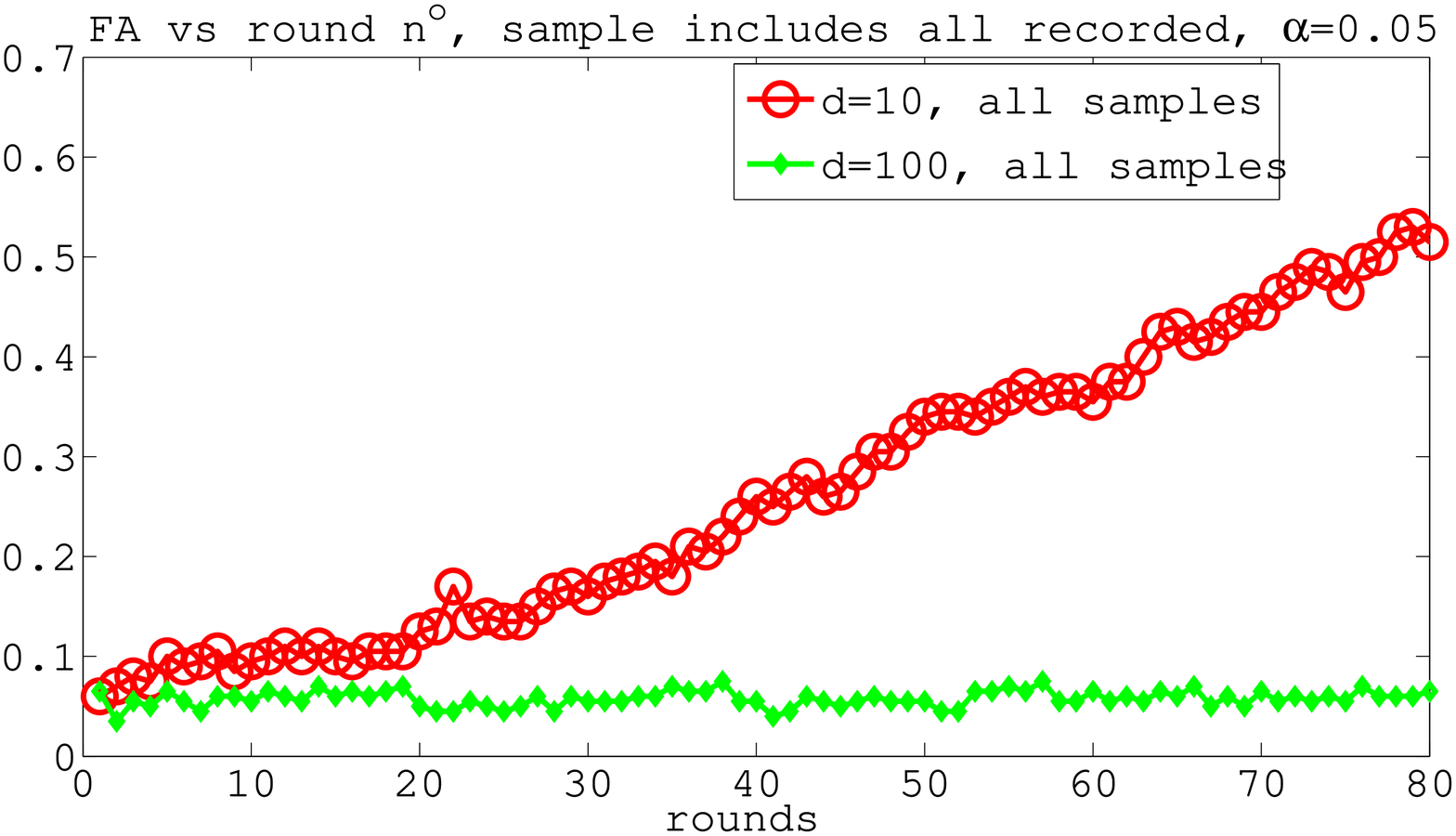, width=2.5in}\\
{\bf (C) - constant mean only for d=100}\\
\epsfig{figure=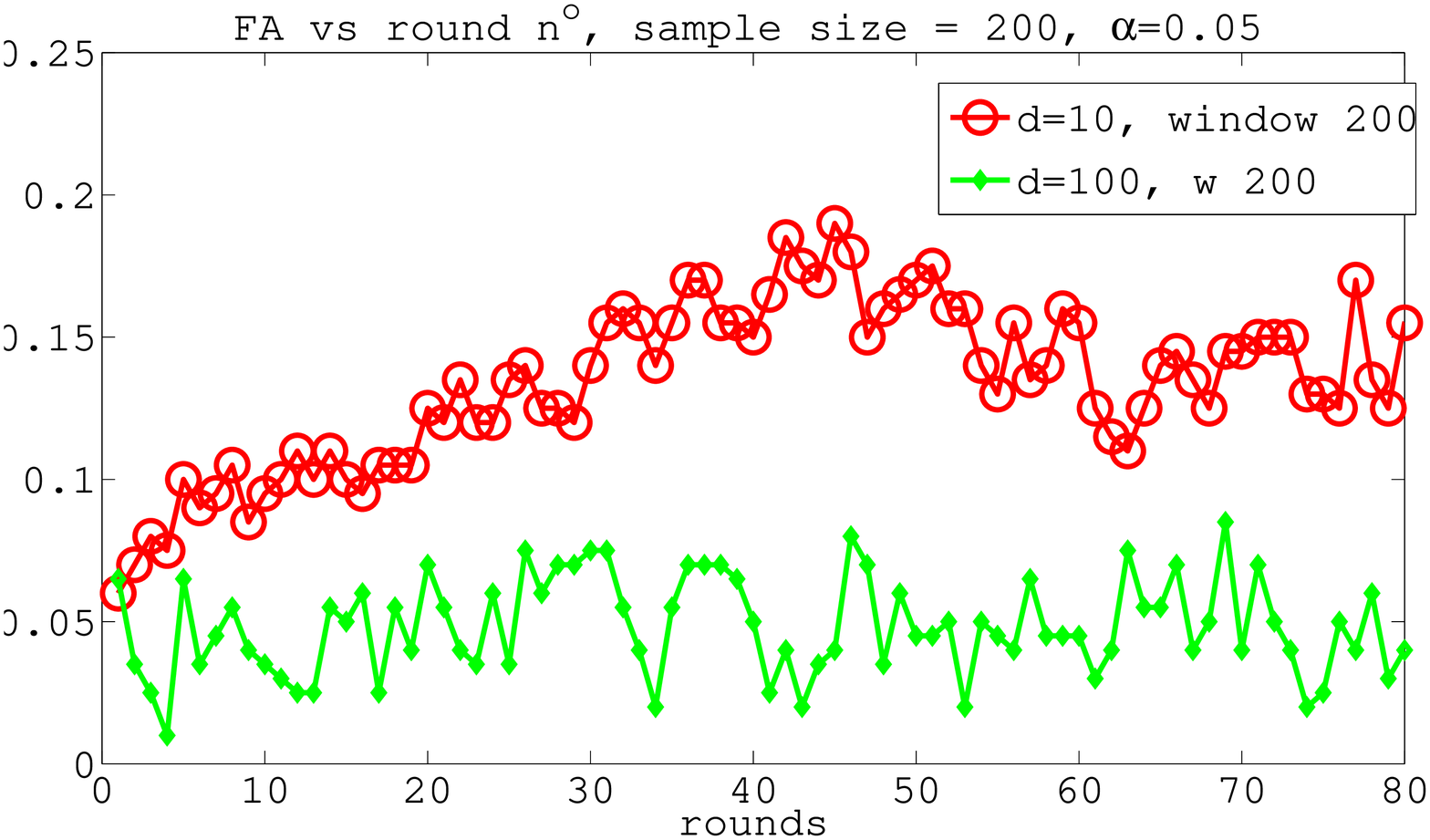 , width=2.5in}\\
{\bf (D) - constant mean only for d=100}
\end{tabular}
\end{center}
\vspace{-0.4cm}
\caption{The effect of the sample size on the False Alarm (FA) trend (FA as a function of the round number, each round containing $d$ samples): {\scriptsize {\bf (A)} For samples drawn from an exponential distribution and {\bf the sample size $di$ at round $i$} FA's mean stays {\em constant}. {\bf (B)} For uniformly sampled transmission times, {\bf the sample size equal to $d$} includes samples mostly from a single round, hence, exponential samples, and FA's mean is constant as in (A) {\bf (C)} For uniform sampling and {\bf the sample size $di$ at round $i$} for $d=100$ FA achieves close approximation with exponential distribution across the rounds, as opposed to $d=10$ that does so only within one round {\bf (D)} At each new round we test 200 preceding samples}} \vspace{-0.8cm} \label{fig:samplesize}
\end{figure}

If $d$ is not sufficiently large, goodness-of-fit tests of exponentiality \cite{AndersonDarling} will fail in a much larger percentage of cases than the value of the test's FA  parameter. For small $d,$ the test results are also sensitive to the sample size as shown in Figure~\ref{fig:samplesize} which compares the pure exponential samples whose false alarm statistics are independent of the sample size (pane {\bf(A)}), with the samples obtained using our group algorithm for  $d=10$ and  $d=100.$ Samples that span just one round will most likely pass the test since ordered uniform samples in any range $\set{(i-1)\mu, i\mu}$ produce intervals described by the exponential distribution $\zeta_{\mu/d}$ (Figure~\ref{fig:samplesize}~{\bf (B)}). Large number of samples would include many rounds, with $1/d$ portion of samples not belonging to the exponential distribution.
Note that for sample size $d=100$ our algorithm achieves close approximation of the exponential distribution as it exhibits the constant FA mean across the rounds, as opposed to $d=10$ (panes {\bf(C)} and {\bf(D)}). In pane {\bf (D)}, at each new round we test 200 preceding samples.
The same sample size was used to test the influence the inserted real events had on the percentage of test failures.
As expected, our simulations confirm that the real events are statistically imperceptible for $d>10.$
\vspace{-0.2cm}
\section{Conclusion}\label{sec:Conclude}
\vspace{-0.1cm}
The proposed decentralized implementations of the fake traffic provide desired statistical source anonymity with minimal  overhead and a delay that depends only on the efficiency of packet routing. Simultaneously, they utilize the network resources in a balanced and fair way, and provide flexibility necessary to handle different temporal and spatial profiles of the event process(omitted here due to space constraints). By designing only one parameter, the size of the dummy population $d,$ according to the known statistical characteristic of the observed process we achieve such flexibility. 
The minimal value of $d$ depends on the implementation, as the deployment of group implementation requires $d$ to be at least $\frac{1}{FA}.$ Uniform spatial distribution of events does not call for the group implementation, and this constraint on $d$ does not hold. However, $d$ needs to be large enough to render $\zeta_{\mu/d}$ and $\zeta_{\mu/(d+1)}$ statistically indistinguishable.

Our future work is to formalize a metric for the quality of a WSN source anonymity scheme that includes the Eve's outage probability and her work needed to collect statistically relevant samples. By including the adversary's work and vulnerabilities, we aim to better model a global eavesdropper, and to present the quality of the source anonymity protection as relative to the adversary's strength. In addition, the quality metric should include the statistically guaranteed anonymity level, the work spent to obfuscate the events, and the latency guarantees by the proposed algorithm.
%
\vspace{-0.2cm}
\bibliographystyle{plain}
\bibliography{EveBib}
\end{document}